\documentclass[10pt]{article}
\usepackage{amsmath}
\usepackage{graphicx}
\usepackage{float}
\usepackage{hyperref}
\usepackage{ragged2e}
\usepackage{amsbsy}
\usepackage{breqn}
\usepackage{amsmath}
\usepackage{amssymb}
\usepackage{empheq}
\usepackage{mathrsfs}
\usepackage{siunitx}
\usepackage{ragged2e} 
\usepackage{etoolbox}
\usepackage{authblk}
\AtBeginEnvironment{align}{\setcounter{subeqn}{0}}
\newcounter{subeqn} \renewcommand{\thesubeqn}{\theequation\alph{subeqn}}%
\newcommand{\subeqn}{
\refstepcounter{subeqn}		
\tag{\thesubeqn}}			

\usepackage[a4paper,bottom=20mm, left=19.05mm, right=19.05mm, top=20mm]{geometry}
\usepackage[utf8]{inputenc}
\usepackage[english]{babel}
\usepackage[final]{changes}  
\setremarkmarkup{(#2)}

\title{Distance Descending Ordering Method: an $O(n)$ Algorithm for Inverting the Mass Matrix in Simulation of Macromolecules with Long Branches}
\author{Xiankun Xu}
\author{Peiwen Li\thanks{corresponding author: peiwen@email.arizona.edu}}
\affil{\footnotesize Department of Aerospace and Mechanical Engineering, The University of Arizona, Tucson, AZ 85721, USA}

\date{\vspace{-6ex}}
\begin{document}
\maketitle
\pagenumbering{arabic}
\justify

\begin{abstract}
Fixman's work in 1974 and the follow-up studies have developed a method that can factorize the inverse of mass matrix into an arithmetic combination of three sparse matrices---one of them is positive definite and need to be further factorized by using the Cholesky decomposition or similar methods. When the molecule subjected to study is of serial chain structure, this method can achieve $O(n)$ computational complexity. However, for molecules with long branches, Cholesky decomposition about the corresponding positive definite matrix will introduce massive fill-in due to its nonzero structure, which makes the calculation in scaling of $O(n^3)$. Although several methods have been used in factorizing the positive definite sparse matrices, no one could strictly guarantee for no fill-in for all molecules according to our test, and thus $O(n)$ efficiency cannot be obtained by using these traditional methods. In this paper we present a new method that can guarantee for no fill-in in doing the Cholesky decomposition, and as a result, the inverting of mass matrix will remain the $O(n)$ scaling, no matter the molecule structure has long branches or not. 
\end{abstract}


\section{Introduction}
\deleted{There are two aspects about the computational scaling, one is the inter-molecular force calculation, such as the calculation of short range force and long rang forces. Another is the inverting of mass matrix.}  

Molecular dynamics simulation involves the calculation of the following equation of motion: $\dot{\mathbf{q}}=\pmb{\mathcal{M}}^{-1}\mathbf{p}$. Where $\mathbf{q}$ is the generalized coordinate, $\mathbf{p}$ is the conjugate momenta, and $\pmb{\mathcal{M}}$ is the mass matrix of the molecule. In Cartesian coordinate the mass matrix is diagonal and is easy to be inverted. For molecule models with geometrical constraints such as fixed angles and lengths, consideration of the molecular movements in the space of generalized internal coordinate rather than Cartesian coordinate is an efficient way to deal with the constraints. In this instance the fixed internal variables are directly excluded from the generalized coordinate, $\mathbf{q}$, and the calculation of constraint forces \cite{ryckaert1977numerical, van1982effect, mazur1997quasi} can be avoided. However, in this case the corresponding mass matrix is a dense matrix. For macromolecules such as DNA, RNA, proteins, polymers, etc., the inverting of mass matrix is a difficult job, the direct calculation of $\pmb{\mathcal{M}}$ and $\pmb{\mathcal{M}}^{-1}$ scaled as $O(n^3)$, where $n$ is the number of atoms in the molecule.

In 1974, Fixman \cite{fixman1974classical} developed a method for efficient calculation of the determinant of mass matrix for polymer chains. Lee et al. \cite{lee2005new, wang2005new, lee2007n} extended this method to solve the equations of motion in $O(n)$ calculation.  They applied Fixman's theorem to invert the mass matrix for structures of serial chain such as planar N-link manipulator, polymer chains, polypeptide chain, and serial chain composed of rigid bodies. They found that the mass matrix inverse can be factorized as $\pmb{\mathcal{M}}^{-1}=\mathbf{A-BC}^{-1}\mathbf{B}^{\top}$, where $\mathbf{A, B}$ and $\mathbf{C}$ are sparse matrices, the number of nonzero entries in the three matrices are all of $O(n)$, also, $\mathbf{C}$ is positive definite. The solving process of the equation of motion can then be decomposed into a series of calculations involving the three sparse matrices. However, a Cholesky decomposition $\mathbf{C}=\mathbf{LL^{\top}}$ is required due to the existence of $\mathbf{C}^{-1}$. If the molecule has serial chain structure, the nonzero entries of $\mathbf{C}$ will be clustered about its diagonal. In this case the calculation of Cholesky decomposition can be of order $O(n)$.  The equation of motion can therefore be solved in an $O(n)$ calculation.

While the above approaches are confined to serial chain structures, in this case the diagonal clustering property of $\mathbf{C}$ makes the Cholesky decomposition easy and efficient. However, if the molecule structures have long branches, the nonzero entries will no longer cluster to the diagonal, even though $\mathbf{C}$ is still sparse. As a result, direct Cholesky decomposition from $\mathbf{C}$ to $\mathbf{L}$ will introduce a lot of fill-in. This means that $\mathbf{L}$ is no longer sparse. The corresponding calculation operation will no longer be linearly scaled with $n$,  but turns to be $O(n^3)$. Therefore, it is necessary to develop a new method to make efficient simulation of macromolecules with long branches.

There are two major categories of numerical methods in dealing with sparse matrices problem: iterative and direct methods. Iterative methods \cite{saad2003iterative, young2014iterative} start with an initial approximation of the solution, and the iteration continues until the desired accuracy being achieved. Typically the number of iterations grows with the size of the matrix. On the other hand, direct methods \cite{george1981computer} obtain the solution in a finite and fixed number of steps. Generally, iterative methods give approximate solutions, while direct methods give solutions up to the machine precision. A typical direct method often associates with finding a permutation matrix $\mathbf{P}$, and applying the Cholesky decomposition on the permuted $\mathbf{PCP^{\top}}$, the target is to make the corresponding $\mathbf{L}$ being as sparse as possible. Common methods in this category include band and envelope methods \cite{cuthill1969reducing, cuthill1972several, felippa1975solution},  minimum degree method \cite{markowitz1957elimination, tinney1967direct, rose1972graph}. Some of these methods are very efficient in reducing the nonzero fills; however, according to our test, none of them can guarantee no fill-in for all types of molecule structure. Also, the procedure of finding the permutation $\mathbf{P}$ has computational load larger than $O(n)$.

The physical aspect of the nonzero structure of $\mathbf{C}$ shows that there are correlations between the nonzero entries and the geometrical structure of the molecule. Based on this observation we proposed a new method to efficiently factorize $\mathbf{C}$, which is especially useful for long branch structures. This method is named as {\em distance descending ordering method}, for it is strongly associated with the distances between atoms. This method has the following features: 1) has $O(n)$ magnitude of calculation; 2) absolutely no fill-in in doing Cholesky decomposition; 3) suitable for molecules with complex structures such as branches, loops, etc., and also, constraints can be arbitrarily applied on angles and lengths; 4) simple, can easily be implemented with computer codes.  

Although we will not discuss the detailed treatment of loop structure in this paper, a macromolecule in general contains not only branches, but also loop structures. The loop can be broken by simply cutting apart one bond inside the loop, and then the loop can be treated as two branches. By doing so we need to deal with the potentials associated with the broken bond. This will only have influence on the calculation of intramolecular forces, and the procedure of inverting mass matrix described in this paper will not be affected.  It is expected, the distance descending ordering method can be applied to molecules with complex structures having both branches and loops.

In section \ref{fixmantheorem}, we will make a brief introduction about the Fixman's theorem and elaborate the difficulties in applying it on molecules with long branches. In section \ref{ddom}, the  distance descending ordering method will be presented with detailed proof. The mechanism why this method has no fill-in in doing the Cholesky decomposition will be explained. In section \ref{numerical}, we apply the method on molecules with 100, 1000 and 10000 atoms to show the no fill-in property. In addition, the $O(n)$ computational efficiency is demonstrated by giving the average calculation time versus the molecule's number of atoms. Finally, in appendix we provide a complement to existing methods \cite{allen1989computer} about calculation of the relative change of internal coordinates with respect to the atoms' position vectors, which is used to calculate the entries of $\mathbf{A}$, $\mathbf{B}$ and $\mathbf{C}$.

\section{Fixman's Theorem}
\label{fixmantheorem}
For a molecule composed by $n+1$ atoms, the atoms are numbered from $0$ to $n$. Where the $0$-th atom is the base atom which we can choose arbitrarily. The position information of this molecule can either be represented by using the Cartesian coordinate as $\mathbf{r}=\{ \vec{r}_0,\vec{r}_1,\cdots,\vec{r}_n\}$, or by the generalized internal coordinate as $\mathbf{g}=\{\vec{g}_0,\vec{g}_1,\cdots,\vec{g}_n\}$, where $\vec{r}_i=\{x_i,y_i,z_i\}$, $\vec{g}_0=\{x_0,y_0,z_0\}$, $\vec{g}_i=\{\phi_i,\theta_i,b_i\}$, and $\phi_i, \theta_i$ and $b_i$ are the torsional angle, bond angle and bond length corresponding to the $i$-th atom, respectively. Since some internal coordinate variables are constrained as constants, the internal coordinate vector $\mathbf{g}$ is partitioned into two parts as $\mathbf{g}=\{\mathbf{q};\mathbf{c}\}$: the soft variables $\mathbf{q}=\{q_1,q_2,\cdots, q_f\}$ and the hard variables $\mathbf{c}=\{c_1,c_2,\cdots,c_r\}$, where $f$ is the degrees of freedom of the molecule, $r$ is the number of constraints, and $f+r=3(n+1)$. Generally $f$ and $r$ are of the same magnitude as $n$ for real molecules, thus there is $O(n)=O(f)=O(r)$.

Since $\mathbf{r}$ and $\mathbf{g}$ are of the same length, the two square Jacobians $\frac{\partial \mathbf{r}}{\partial \mathbf{g}}$ and $\frac{\partial \mathbf{g}}{\partial \mathbf{r}}$ are inverse of each other:
\begin{equation}
	\left(\frac{\partial \mathbf{r}}{\partial \mathbf{g}}\right)^{-1} =	\frac{\partial \mathbf{g}}{\partial \mathbf{r}}
\label{eqA}
\end{equation}
where 
\begin{equation}
\begin{split}
	& \frac{\partial \mathbf{r}}{\partial \mathbf{g}} = \left[\left.\frac{\partial \mathbf{r}}{\partial \mathbf{q}} \right \vert \frac{\partial \mathbf{r}}{\partial \mathbf{c}}\right] 
	=\left[\left.
	\begin{matrix}
		\left[\frac{\partial \vec{r}_0}{\partial q_1} \right] &\left[\frac{\partial \vec{r}_0}{\partial q_2} \right] & \cdots &\left[\frac{\partial \vec{r}_0}{\partial q_f} \right]\\
		\vdots & \vdots & \cdots & \vdots  \\
		\left[\frac{\partial \vec{r}_n}{\partial q_1} \right] &\left[\frac{\partial \vec{r}_n}{\partial q_2} \right] & \cdots &\left[\frac{\partial \vec{r}_n}{\partial q_f} \right]	
	\end{matrix}
	 \right \vert 
	 \begin{matrix}
 		\left[\frac{\partial \vec{r}_0}{\partial c_1} \right] &\left[\frac{\partial \vec{r}_0}{\partial c_2} \right] & \cdots &\left[\frac{\partial \vec{r}_0}{\partial c_r} \right]\\
 		\vdots & \vdots & \cdots & \vdots  \\
 		\left[\frac{\partial \vec{r}_n}{\partial c_1} \right] &\left[\frac{\partial \vec{r}_n}{\partial c_2} \right] & \cdots &\left[\frac{\partial \vec{r}_n}{\partial c_r} \right]	
     \end{matrix}
	 \right] 	\\
	&\frac{\partial \mathbf{g}}{\partial \mathbf{r}} =
		\left[
		\begin{array}{c}
		\partial \mathbf{q} / \partial \mathbf{r} \\
		\hline
		\partial \mathbf{c} / \partial \mathbf{r}
		\end{array}
		\right] =
		\left[
		\begin{array}{c c }
		\begin{matrix}
			\left[ \partial  q_1/ \partial \vec{r}_0 \right] &\left[ \partial  q_1/ \partial \vec{r}_1 \right] & \cdots &\left[ \partial  q_1/ \partial \vec{r}_n \right]\\
			\vdots & \vdots & \cdots & \vdots  \\
			\left[ \partial  q_f/ \partial \vec{r}_0 \right] &\left[ \partial  q_f/ \partial \vec{r}_1 \right] & \cdots &\left[ \partial  q_f/ \partial \vec{r}_n \right]
		\end{matrix} \\
		\hline
		\begin{matrix}
			\left[ \partial  c_1/ \partial \vec{r}_0 \right] &\left[ \partial  c_1/ \partial \vec{r}_1 \right] & \cdots &\left[ \partial  c_1/ \partial \vec{r}_n \right]\\
			\vdots & \vdots & \cdots & \vdots  \\
			\left[ \partial  c_r/ \partial \vec{r}_0 \right] &\left[ \partial  c_r/ \partial \vec{r}_1 \right] & \cdots &\left[ \partial  c_r/ \partial \vec{r}_n \right]
		\end{matrix}
		\end{array}
		\right] 
\end{split}
\label{eqB}
\end{equation}

In the above equations, $\frac{\partial \mathbf{r}}{\partial \mathbf{g}}$ is a dense matrix, while $\frac{\partial \mathbf{g}}{\partial \mathbf{r}}$ is a very sparse matrix. It can be seen in Eq. \eqref{eqC} in appendix that $\phi_i, \theta_i$ and $b_i$ are only affected by 4, 3 and 2 atoms, respectively, and thus most of the entries in $\frac{\partial \mathbf{g}}{\partial \mathbf{r}}$ are zero. We want to use Eq. \eqref{eqA} together with the sparse property of $\frac{\partial \mathbf{g}}{\partial \mathbf{r}}$ to efficiently invert the mass matrix.

Define matrix $\mathbf{G}$ as follows:
\begin{equation}
\begin{split}
	\mathbf{G} &=\left(\frac{\partial \mathbf{r}}{\partial \mathbf{g}}\right)^{\top} \mathbf{M}\left(\frac{\partial \mathbf{r}}{\partial \mathbf{g}}\right) \\
	&= \left[
	 \begin{array}{c}
		 \left(\partial \mathbf{r} / \partial \mathbf{q}\right)^{\top} \\
		 \hline
		 \left(\partial \mathbf{r} / \partial \mathbf{c}\right)^{\top}
	 \end{array}
	\right] 
	\left[
		\begin{matrix}
			m_0 \mathbf{I}_3\\
			  & \ddots  \\
			  & &  m_n \mathbf{I}_3  		
		\end{matrix}
	\right]
	\left[\left.\frac{\partial \mathbf{r}}{\partial \mathbf{q}} \right \vert \frac{\partial \mathbf{r}}{\partial \mathbf{c}}\right] \\
	&= \left[
		 \begin{array}{c|c}
			 \left(\frac{\partial \mathbf{r}}{\partial \mathbf{q}}\right)^{\top} \mathbf{M}\left(\frac{\partial \mathbf{r}}{\partial \mathbf{q}}\right) &
			  \left(\frac{\partial \mathbf{r}}{\partial \mathbf{q}}\right)^{\top} \mathbf{M}\left(\frac{\partial \mathbf{r}}{\partial \mathbf{c}}\right)\\
			 \hline
			 \left(\frac{\partial \mathbf{r}}{\partial \mathbf{c}}\right)^{\top} \mathbf{M}\left(\frac{\partial \mathbf{r}}{\partial \mathbf{q}}\right) &
			  \left(\frac{\partial \mathbf{r}}{\partial \mathbf{c}}\right)^{\top} \mathbf{M}\left(\frac{\partial \mathbf{r}}{\partial \mathbf{c}}\right)
		 \end{array}
		\right]
\end{split}
\label{eqD}
\end{equation}
The mass matrix $\pmb{\mathcal{M}}=\left(\frac{\partial \mathbf{r}}{\partial \mathbf{q}}\right)^{\top} \mathbf{M}\left(\frac{\partial \mathbf{r}}{\partial \mathbf{q}}\right) $ is the first block of $\mathbf{G}$. From Eq. \eqref{eqA}, we have  
\begin{equation}
\begin{split}
	& \mathbf{G}^{-1}=\left(\frac{\partial \mathbf{r}}{\partial \mathbf{g}}\right)^{-1} \mathbf{M}^{-1}\left(\frac{\partial \mathbf{r}}{\partial \mathbf{g}}\right)^{-\top}=
	\left(\frac{\partial \mathbf{g}}{\partial \mathbf{r}}\right) \mathbf{M}^{-1} \left(\frac{\partial \mathbf{g}}{\partial \mathbf{r}}\right)^{\top} \\
	&= \left[
		 \begin{array}{c|c}
			 \left(\frac{\partial \mathbf{q}}{\partial \mathbf{r}}\right) \mathbf{M}^{-1}\left(\frac{\partial \mathbf{q}}{\partial \mathbf{r}}\right)^{\top} &
			  \left(\frac{\partial \mathbf{q}}{\partial \mathbf{r}}\right) \mathbf{M}^{-1}\left(\frac{\partial \mathbf{c}}{\partial \mathbf{r}}\right)^{\top} \\
			 \hline
			 \left(\frac{\partial \mathbf{c}}{\partial \mathbf{r}}\right) \mathbf{M}^{-1}\left(\frac{\partial \mathbf{q}}{\partial \mathbf{r}}\right)^{\top} &
			  \left(\frac{\partial \mathbf{c}}{\partial \mathbf{r}}\right) \mathbf{M}^{-1}\left(\frac{\partial \mathbf{c}}{\partial \mathbf{r}}\right)^{\top}
		 \end{array}
		\right]
	= \left[
	\begin{array}{c|c}
		\mathbf{A} &
		\mathbf{B}\\
		\hline
		\mathbf{B}^{\top} &
		\mathbf{C}
	\end{array}
	\right]
\end{split}
\label{eqD1}
\end{equation}
By comparing $\mathbf{G}$ and $\mathbf{G}^{-1}$ \cite{lee2005new, wang2005new, lee2007n}, we have:
\begin{equation}
\begin{split}
	\pmb{\mathcal{M}}^{-1}=\mathbf{A-BC^{-1}B^{\top}}
\end{split}
\label{eqE}
\end{equation}
The favorable feature is that $\mathbf{A}, \mathbf{B}$ and $\mathbf{C}$ are all sparse matrices, the number of nonzero entries in this three matrices are all of order $O(r)$. Also, both $\mathbf{A}(f \mbox{ by } f)$ and $\mathbf{C}(r \mbox{ by } r)$ are positive definite matrices.  

The inverting of mass matrix problem $\mathbf{\dot{q}}=\pmb{\mathcal{M}}^{-1}\mathbf{p}$ can then be solved as:
\begin{empheq}[left=\empheqlbrace]{align}
	\mathbf{y}_1 & =\mathbf{Ap} \refstepcounter{equation}\subeqn  \label{eqG1}\\
	\mathbf{y}_2 & =\mathbf{B^{\top} p}   \subeqn \label{eqG2}	\\
	\mathbf{y}_3 & =\mathbf{C}^{-1} \mathbf{y}_2   \subeqn \label{eqG3}	\\
	\mathbf{y}_4 & =\mathbf{B}  \mathbf{y}_3    \subeqn	\label{eqG4}\\
	\mathbf{\dot{q}} & = \mathbf{y}_1-\mathbf{y}_4  \subeqn \label{eqG5}
\end{empheq}
where calculation of Eq. \eqref{eqG1} can be further decomposed as $\mathbf{y}_1=\frac{\partial \mathbf{q}}{\partial \mathbf{r}} \cdot \left\{\mathbf{M}^{-1}\cdot \left[\left(\partial \mathbf{q}/\partial \mathbf{r}\right)^{\top} \cdot \mathbf{p} \right] \right\} $, and the calculation of Eqs. \eqref{eqG2}, \eqref{eqG4} are similar. Because the number of nonzero entries in $\mathbf{A}$ and $\mathbf{B}$ are of the order of $O(r)$, the complexity of calculation of these three equations will also have the order of $O(r)$.

\begin{figure}[!h]
\centering		
	\minipage{0.34\textwidth}
	\includegraphics[width=0.89\linewidth] {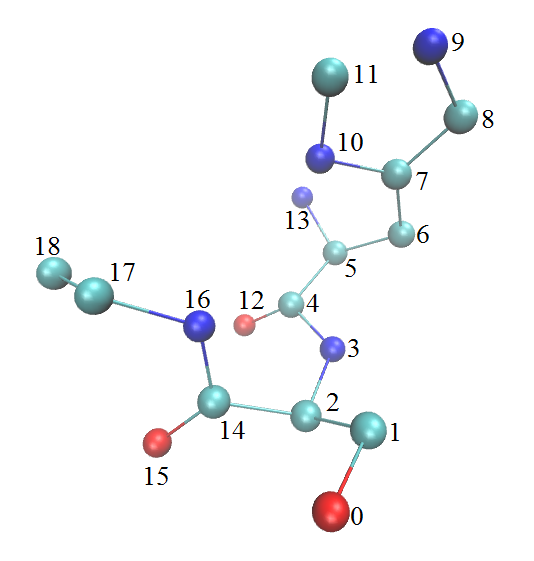} 
	\endminipage		
	\minipage{0.35\textwidth}
	\includegraphics [width=0.88\linewidth]{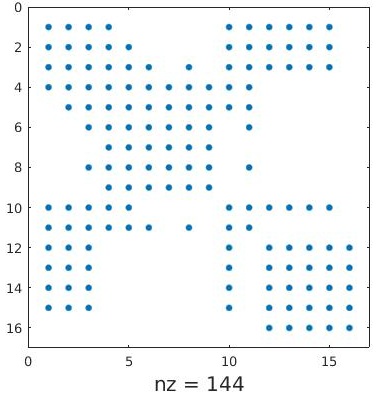}
	\endminipage
	\minipage{0.35\textwidth}
	\includegraphics [width=0.88\linewidth]{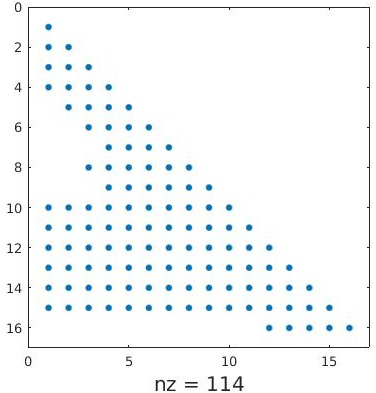}
	\endminipage
	\caption{a numbered molecule (left), the nonzero structures of $\mathbf{C}$ (middle) and $\mathbf{L}$ (right)} 
	\label{glucagon}
\end{figure}

Since $\mathbf{C}$ is positive definite, we can use Cholesky decomposition to factorize it as $\mathbf{C=LL^{\top}}$, and the solving of Eq. \eqref{eqG3} becomes $\mathbf{L}\left(\mathbf{L}^{\top}\mathbf{y}_3\right)=\mathbf{y}_2$. Here the critical issue is that even though $\mathbf{C}$ is a sparse matrix, the corresponding $\mathbf{L}$ may not necessarily be. If a molecule has long branches, the Cholesky decomposition from $\mathbf{C}$ to $\mathbf{L}$ may introduce fill-in, namely, $\mathbf{L}$ has nonzero entries in the positions which are zero in $\mathbf{C}$. Take the molecule shown in Fig. \ref{glucagon} for demonstration, the hard variables are set as $\mathbf{c}=\left\{\phi_3,\phi_4,\cdots,\phi_{18} \right\}$, the nonzero structures of the corresponding $\mathbf{C}$ and $\mathbf{L}$ are also shown. It could be found that $\mathbf{L}$ introduces some fill-in when compared to $\mathbf{C}$. As a result, the computation magnitude of the Cholesky decomposition will approach to $O(r^3)$ and the solving of $\mathbf{L}\left(\mathbf{L}^{\top}\mathbf{y}_3\right)=\mathbf{y}_2$ becomes $O(r^2)$, which is certainly unacceptable when $r$ becomes very large. 

\section{Descending Distance Ordering Method}
\label{ddom}
The most optimum condition is that no fill-in will be introduced in the Cholesky decomposition, in this situation $\mathbf{L}$ has the same nonzero structure as that of the lower diagonal part of $\mathbf{C}$. In this section, the {\em distance descending ordering method} will be proposed, it can strictly guarantee no fill-in, and thus gives the $O(r)$ computational efficiency. Unlike most of the direct methods in dealing with sparse matrices, this method does not need to explicitly do row and column permutations to find $\mathbf{P}$, instead it directly gives an ordering strategy about the hard variable list $\mathbf{c}$, based on the structure of the molecule.  It is especially powerful for molecules with long branches. In addition, geometrical constraints can be arbitrarily applied. This method is also very simple, makes it easy to be integrated into computer codes.

In the following, we start the discussion with 2 definitions:

\noindent \textbf{Correlation}: For the matrix $\mathbf{C}$ defined in Eq. \eqref{eqD1}, if its $ij$-th entry is nonzero, we call that the two hard variables $c_i$ and $c_j$ are correlated, or $c_i$ has correlation with $c_j$, where $c_i$ and $c_j$ are the $i$-th and $j$-th elements in hard variable vector $\mathbf{c}$.

\noindent \textbf{Distance}: The distance of the $i$-th atom is defined as the number of bonds between itself and the $0$-th atom, which has a unique value if there is no loops.

The following three points about the structure of $\mathbf{C, L}$ and the permutation operation should be noted:
\begin{enumerate}
\item
	The number of nonzero entries in matrix $\mathbf{C}$ is of order $O(r)$, since each hard variable $c_i$ only correlates with very few number of other hard variables. In other words, the nonzero entries in each line or column is finite and is the order of $O(10)$. This is because $\mathbf{C}_{ij}=\sum_{k=0}^{n}\frac{1}{m_k}\frac{\partial c_i}{\partial \vec{r}_k}\cdot \frac{\partial c_j}{\partial \vec{r}_k} \neq 0$ requires both $c_i$ and $c_j$ being affected by at least one same $\vec{r}_k$, and from Eq. \eqref{eqC} we see that the internal variables are only related to atoms that are close or nearby.
\item
	If no fill-in, the number of nonzero entries in matrix $\mathbf{L}$ is of the order of $O(r)$. Also, the computational complexity in applying Cholesky decomposition and calculation of $\mathbf{L}\left(\mathbf{L}^{\top}\mathbf{y}_3\right)=\mathbf{y}_2$ are both of the order of $O(r)$.
\item
	A permutation on matrix $\mathbf{C}$ is actually a permutation on the hard variable list $\mathbf{c}$, or $\mathbf{P}$ applied a reordering operation on vector $\mathbf{c}$, which could be seen as follows:
	\begin{equation}
	\begin{split}
		\mathbf{PCP^{\top}} & = \mathbf{P} \left(\frac{\partial \mathbf{c}}{\partial \mathbf{r}}\right) \mathbf{M}^{-1}\left(\frac{\partial \mathbf{c}}{\partial \mathbf{r}}\right)^{\top} \mathbf{P}^{\top}=\left(\frac{\partial \mathbf{Pc}}{\partial \mathbf{r}}\right) \mathbf{M}^{-1}\left(\frac{\partial \mathbf{Pc}}{\partial \mathbf{r}}\right) \\
		&= \left(\frac{\partial \mathbf{c'}}{\partial \mathbf{r}}\right) \mathbf{M}^{-1}\left(\frac{\partial \mathbf{c'}}{\partial \mathbf{r}}\right) =\mathbf{LL^{\top}}
	\end{split}
	\label{eqH}
	\end{equation}	 
	
\end{enumerate}
where $\mathbf{c'}=\mathbf{Pc}$ is the reordered hard variable vector. Instead of directly permute matrix $\mathbf{C}$, we are going to find an ordering strategy about hard variable vector $\mathbf{c}$, to make sure that there is no fill-in in the procedure of Cholesky decomposition. 

\deleted{We now relate the fill-in to the correlations between hard variables. }Consider a basic factorization step of Cholesky decomposition:
\begin{equation}
\begin{split}
	\mathbf{C}_r
	& =\left(
	\begin{matrix}
	C_{11} & \mathbf{v}_1^{\top}\\
	\mathbf{v}_1 &  \mathbf{C}_{r-1}  		
	\end{matrix}
	\right) 
	=\left(
	\begin{matrix}
	\sqrt{C_{11}} & \mathbf{0} \\
	\frac{\mathbf{v}_1}{\sqrt{C_{11}}}  &  \mathbf{I}_{r-1}
	\end{matrix}
	\right) 
	\left(
	\begin{matrix}
	1 & \mathbf{0} \\
	\mathbf{0}  &  \mathbf{C}_{r-1}-\frac{\mathbf{v}_1 \mathbf{v}_1^{\top}}{C_{11}}
	\end{matrix}
	\right) 
	\left(
	\begin{matrix}
	\sqrt{C_{11}} & \frac{\mathbf{v}_1^{\top}}{\sqrt{C_{11}}} \\
	\mathbf{0}  &  \mathbf{I}_{r-1}
	\end{matrix}
	\right)  \\
	& =\mathbf{L}_1 
	\left(
	\begin{matrix}
	1 & \mathbf{0} \\
	\mathbf{0}  &  \mathbf{C}_{r-1}-\frac{\mathbf{v}_1 \mathbf{v}_1^{\top}}{C_{11}}
	\end{matrix}
	\right) \mathbf{L}_1^{\top}
\end{split}
\end{equation}
Set $\mathbf{C}_{r-1}^'=\mathbf{C}_{r-1}-\mathbf{v}_1\mathbf{v}_1^{\top}/\mathbf{C}_{11}$, $\mathbf{C}_{r-1}^'$ is positive definite and the above basic step is recursively applied. The no fill-in requires that $\mathbf{C}_{r-1}^'$ has the same nonzero structure as $\mathbf{C}_{r-1}$. Three observations can be made about the factorization step:
\begin{enumerate}
\item
	For each nonzero entry of $\mathbf{v}_1\mathbf{v}_1^{\top}/\mathbf{C}_{11}$, if the corresponding entry of $\mathbf{C}_{r-1}$ is also nonzero, then $\mathbf{C}_{r-1}^'$ will have the same nonzero structure as $\mathbf{C}_{r-1}$.
\item
	The $ij$-th entry in $\mathbf{v}_1\mathbf{v}_1^{\top}/\mathbf{C}_{11}$ is nonzero, which requires the two hard variables $c_i$ and $c_j$ being both correlated with the eliminating hard variable $c_1$. 
\item 
	The $ij$-th entry in $\mathbf{C}$ is nonzero, which implies that the two hard variables $c_i$ and $c_j$ are correlated. 
\end{enumerate}
Therefore, in order to guarantee for no fill-in, the hard variables which have correlation with the eliminating hard variable should be mutually correlated with each other.

\begin{figure}[!h]
\centering
	\includegraphics[width=0.99\linewidth]{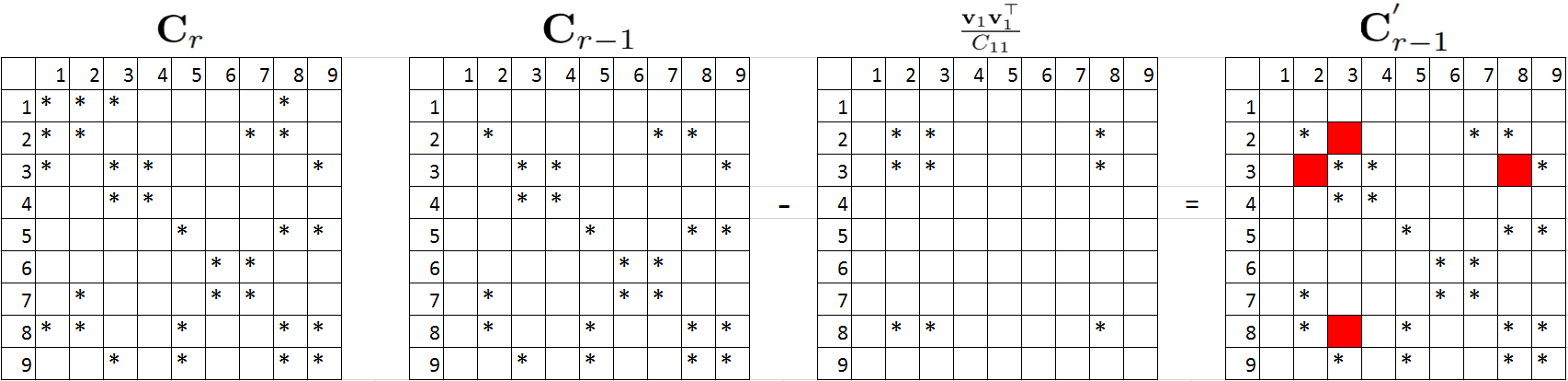}
	\caption{Correlation determines fill-in.}
	\label{correlation}
\end{figure}

This statement can be demonstrated by Fig. \ref{correlation}, where hard variables $c_2, c_3$ and $c_8$ are correlated with $c_1$. The requirement of no fill-in demands that $c_2, c_3$ and $c_8$ should be mutually correlated with each other. However, $c_3$ is not correlated with $c_2$ and $c_8$, nonzero entries are introduced in the $(2,3)$, $(3,8)$ entries as well as their symmetric entries.

Thus the correlation between hard variables plays a critical role  in determining if fill-in would happen or not. Three points may be noted about the correlation between hard variables:
\begin{enumerate}
\item
	The correlation between $c_i$ and $c_j$ implies that these two hard variables are affected by at least one identical atom.
\item
	Different types of hard variables are affected by different number of atoms. As shown in Eq. \eqref{eqC}, $\phi_i, \theta_i$ and $b_i$ are affected by 4, 3 and 2 atoms, respectively.
\item
	Hard variables in different branches may also have correlation. For the molecule shown in Fig. \ref{glucagon}, $\phi_4$ and $\phi_{16}$ are correlated since both are affected by the $1$st and $2$nd atoms. And this is the reason long branches may introduce fill-in.
\end{enumerate}

Point 2 suggested that the correlation depends on the types of the hard variables. To make the discussion clear, in Sec. 3.1 we will propose the distance descending ordering method for the case that only torsional angles are allowed to be hard variable. In this case each hard variable is affected by 4 atoms, making the discussion be simplified. In Sec. 3.2, the method is extended to allow all internal variables being arbitrarily constrained. Indeed, the first case is a special situation of the second one.

\subsection{Only Torsional Angles are Allowed to be Hard Variable}
\noindent \textbf{Theorem 1}: In the case that only torsional angles are allowed to be constrained, if the hard variables in vector $\mathbf{c}$ are arranged in the descending order according to the distances to the $0$-th atom, then the Cholesky decomposition on matrix $\mathbf{C}$ will have no fill-in.

\noindent \textbf{Proof}:
\begin{enumerate}
\item
	If $c_1$ is the eliminating hard variable and it has the largest distance to $0$-th atom, then $c_1$ will only correlate with hard variables which have less or equal distances. The distance differences between $c_1$ and these hard variables are less than or equal to 4 (a limit on the difference of distance if two torsional angles need to be correlated). As a result, the distance differences between these hard variables are also less than or equal to 4, and all these hard variables will mutually correlate with each other. Thus the elimination of $c_1$ guarantees no fill-in. 
\item  
	After the elimination of $c_1$, its corresponding atom can be eliminated from the molecule structure, and its correlations with the remaining hard variables will no longer have influence on the following basic factorization steps . The largest distance torsional angles among the remaining are now considered as the next eliminating hard variables. The above logic is again applied, this guarantees that each basic factorization step has no fill-in.
\end{enumerate}

\begin{figure}[!h]
\centering		
	\minipage{0.5\textwidth}
	\includegraphics[width=0.89\linewidth] {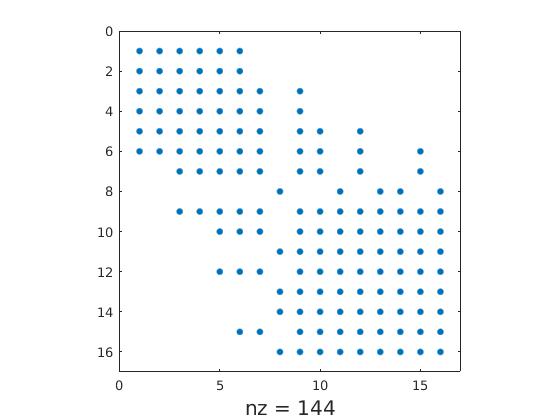} 
	\endminipage		
	\minipage{0.5\textwidth}
	\includegraphics [width=0.89\linewidth]{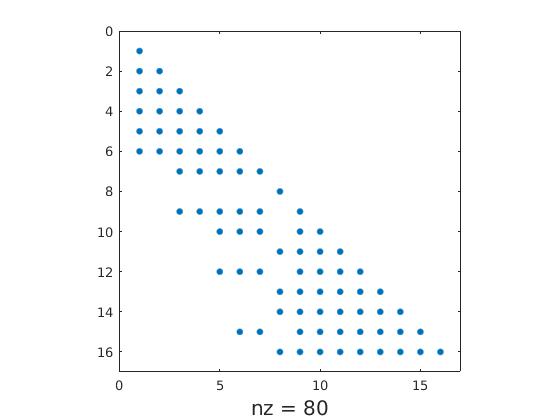}
	\endminipage
	\caption{Distance descending ordering method has no fill-in: $\mathbf{C}$ (left); $\mathbf{L}$ (right)} 
	\label{glucagonOrder}
\end{figure}

Theorem 1 can be best elaborated by using the molecule shown in Fig. \ref{glucagon} as an example. Suppose all torsional angles are constrained except $\phi_1$ and $\phi_2$, then the hard variable vector $\mathbf{c}$ should be ordered as $\mathbf{c}=\{\phi_9, \phi_{11}, \phi_8, \phi_{10}, \phi_7, \phi_6, \phi_{13}, \phi_{18}, \phi_5, \phi_{12}, \phi_{17}, \phi_4, \phi_{15}, \phi_{16}, \phi_3, \phi_{14}\}$. The corresponding $\mathbf{C}$ and $\mathbf{L}$ are shown in Fig. \ref{glucagonOrder}, it is seen that there is no fill-in introduced.

\subsection{Any Internal Coordinate Variable can be Constrained}
In the case of arbitrarily constrained internal variables, because different types of internal coordinate variables are influenced by different number of atoms, the correlation between hard variables varies according to their types. As a result, simply ordering the hard variables in the descending order of distances will introduce nonzero fill-in, since it can not guarantee that the hard variables which have correlations with the eliminating hard variable are mutually correlated with each other.

In fact, the descending ordering strategy should still be applied, but with a slight modification, which can thus avoid the introducing of fill-in:

\noindent \textbf{Theorem 2}: In the case that any internal variables can be considered as hard variable, the hard variables should first be ordered within the distance level $l$ (collection of atoms have distance $l$ to the base atom) as: $[b_1^{(l)}, b_2^{(l)}, \cdots,b_{n_l}^{(l)}$; $\phi_1^{(l+1)}, \phi_2^{(l+1)}, \cdots,\phi_{n_{l+1}}^{(l+1)}$; $\theta_1^{(l)}, \theta_2^{(l)}, \cdots,\theta_{n_l}^{(l)}]$. And then the distance levels are still arranged in a descending order. The corresponding Cholesky decomposition on matrix $\mathbf{C}$ will have no fill-in.

The terms $n_l$ and $n_{l+1}$ are the number of atoms having distances $l$ and $l+1$ to the base atom, the superscripts $(l)$ and $(l+1)$ denote that these hard variables belong to atoms that have distances $l$ and $l+1$, respectively. It is the rule that if any internal variable is non-constant, it should be excluded from the hard variable vector $\mathbf{c}$.

Again, take the molecule shown in Fig. \ref{glucagon} as an example, the hard variables should be ordered as follows: $\mathbf{c}=\{[b_9,b_{11};\theta_9,\theta_{11}]$, $[b_8,b_{10};\phi_9,\phi_{11}; \theta_8,\theta_{10}]$, $\cdots$, $[b_2;\phi_3,\phi_{14};\theta_2], [b_1] \}$, where $x_0, y_0, z_0$ and $\phi_1, \theta_1, \phi_2$ are excluded from $\mathbf{c}$ because they represent the translational and rotational movements of the molecule and must be non-constant. Also, non-constant hard variables should be removed from this list.

This ordering strategy takes into account the fact that $b_i, \phi_i$ and $\theta_i$ are affected by different number of atoms. It can be verified that in applying this ordering method, hard variables which correlate with the eliminating hard variable are also mutually correlated with each other. As a result, the Cholesky decomposition will have no fill-in. 

\section{Numerical Validations}
\label{numerical}
In section 3, we have proven that if no fill-in in Cholesky decomposition about matrix $\mathbf{C}$, then the calculation complexity of Eq. \eqref{eqG3} will be of the order of $O(r)$. Correspondingly, the solving of the equation $\mathbf{\dot{q}}=\pmb{\mathcal{M}}^{-1}\mathbf{p}$ will obtain the $O(r)$ computational efficiency.

\begin{figure}[!h]
\centering		
	\minipage{0.255\textwidth}
	\includegraphics[width=0.89\linewidth] {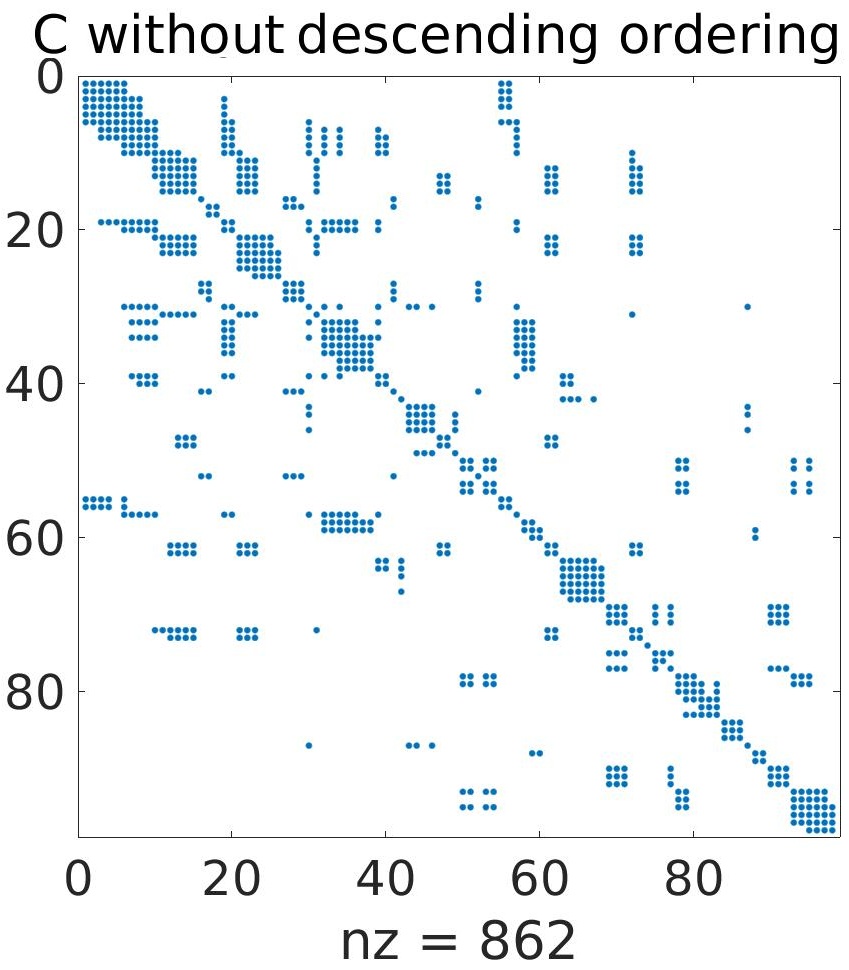} 
	\endminipage		
	\minipage{0.255\textwidth}
	\includegraphics [width=0.89\linewidth]{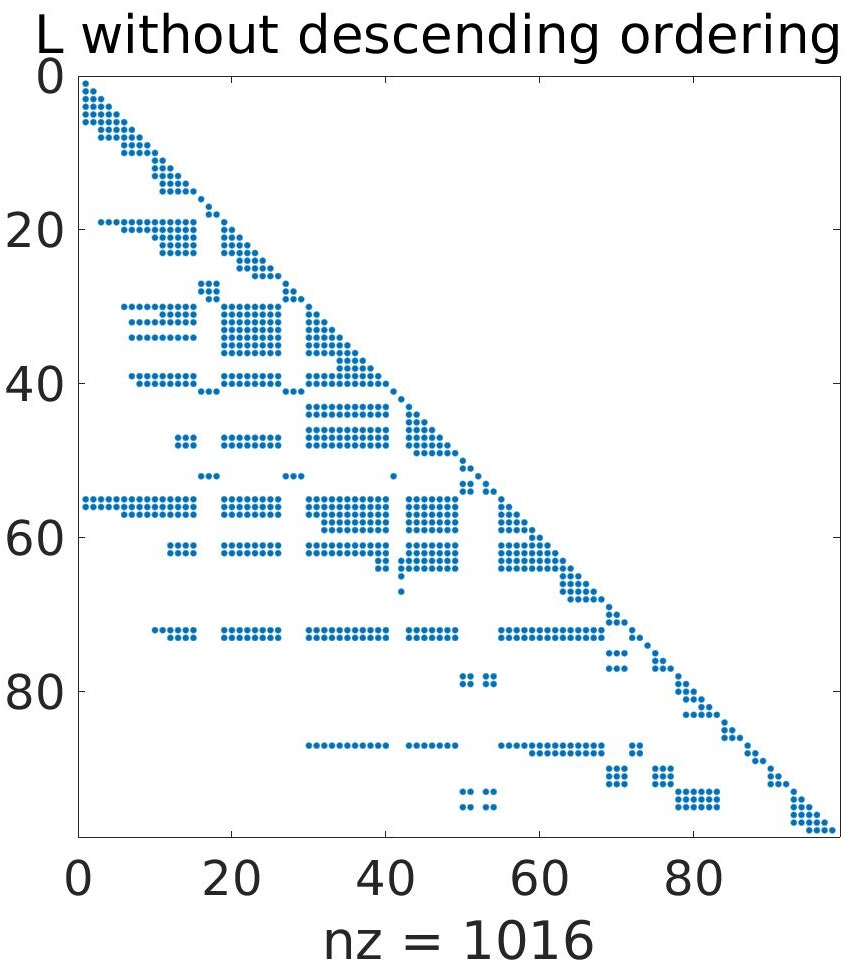}
	\endminipage
	\minipage{0.255\textwidth}
	\includegraphics[width=0.89\linewidth] {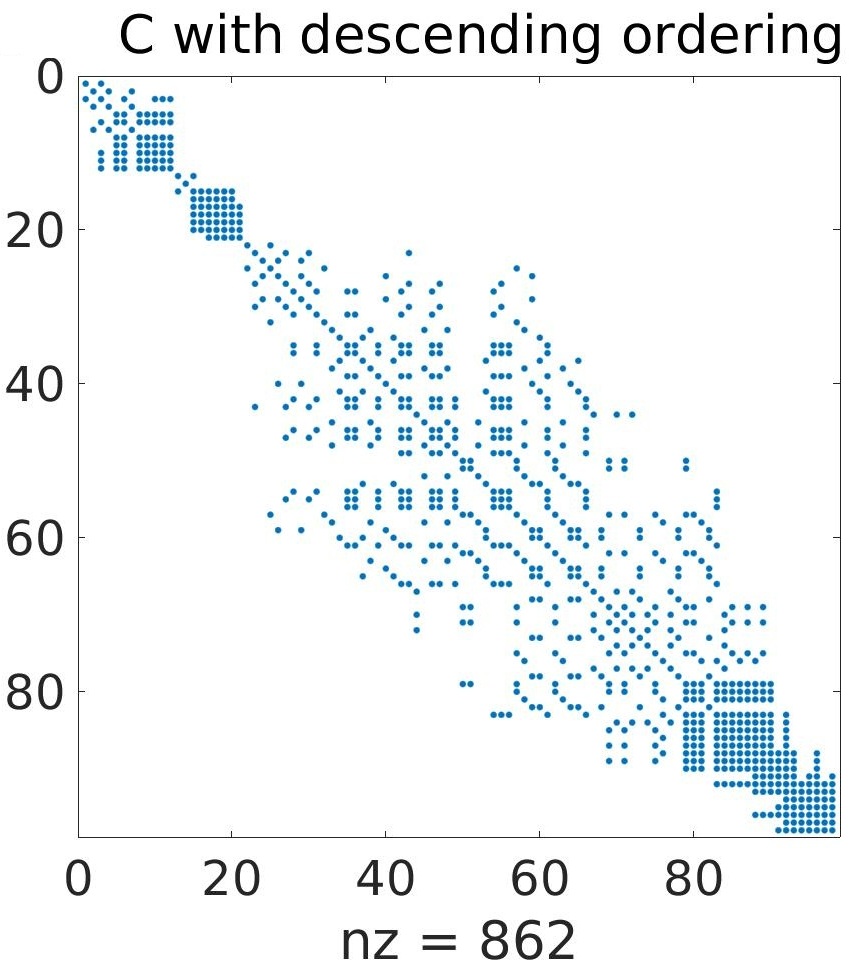} 
	\endminipage		
	\minipage{0.255\textwidth}
	\includegraphics [width=0.89\linewidth]{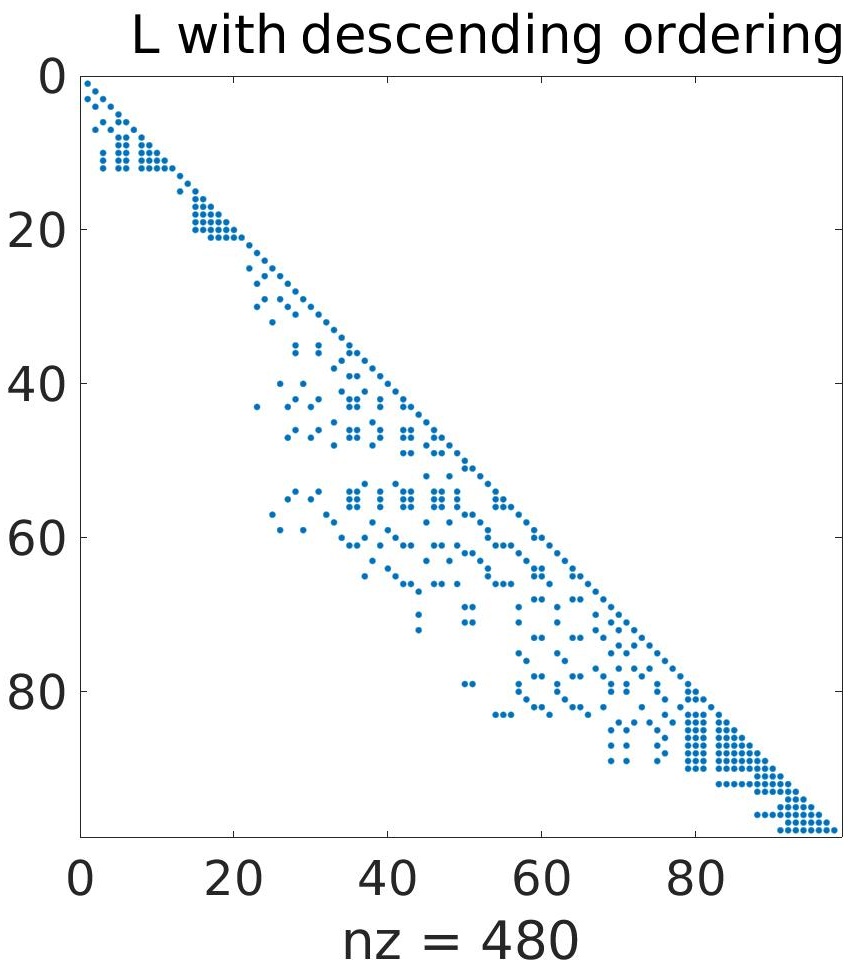}
	\endminipage
	\caption{Nonzero structures of $\mathbf{C}$ and $\mathbf{L}$ from algorithms with and without descending ordering, molecule with $n=100, r=98$.}
	\label{n100}
\end{figure}
\begin{figure}[!h]
\centering		
	\minipage{0.256\textwidth}
	\includegraphics[width=0.89\linewidth] {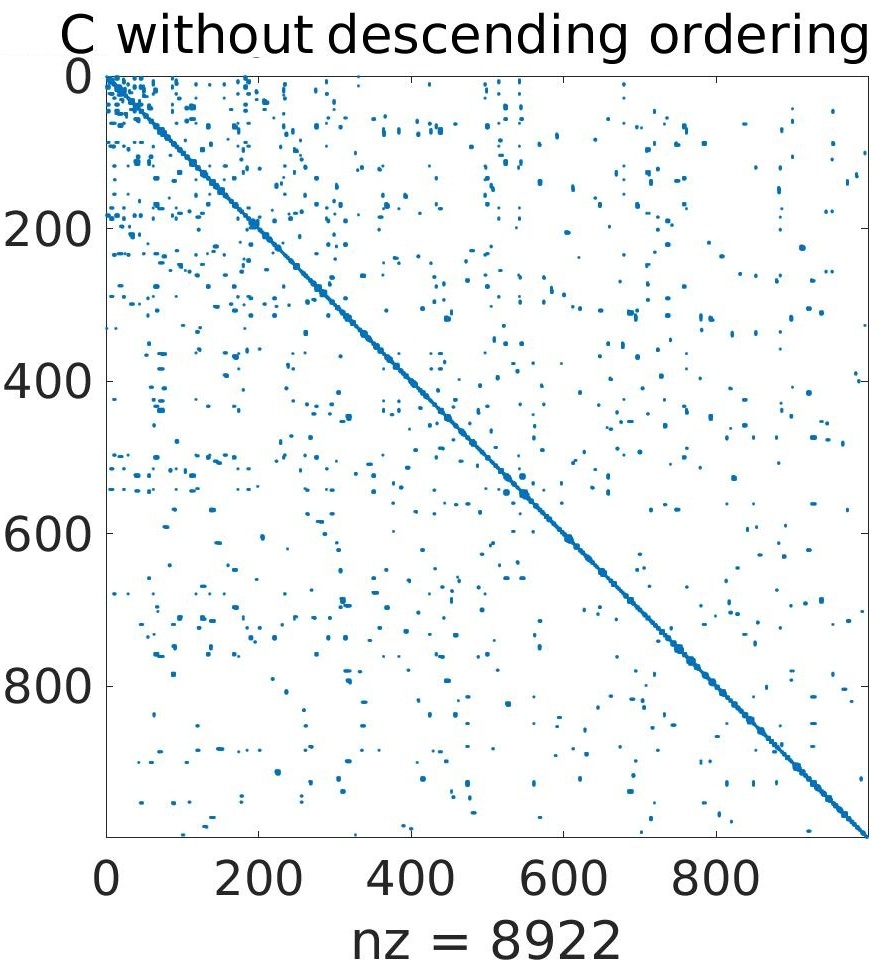} 
	\endminipage		
	\minipage{0.256\textwidth}
	\includegraphics [width=0.89\linewidth]{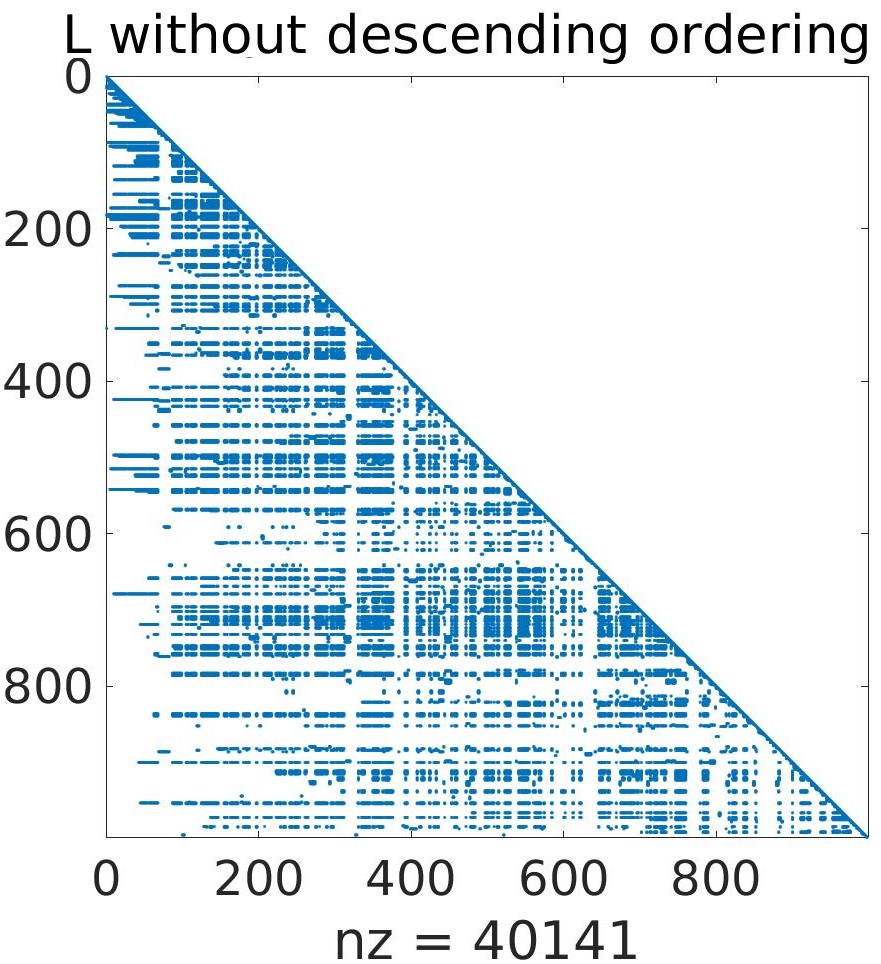}
	\endminipage
	\minipage{0.256\textwidth}
	\includegraphics[width=0.89\linewidth] {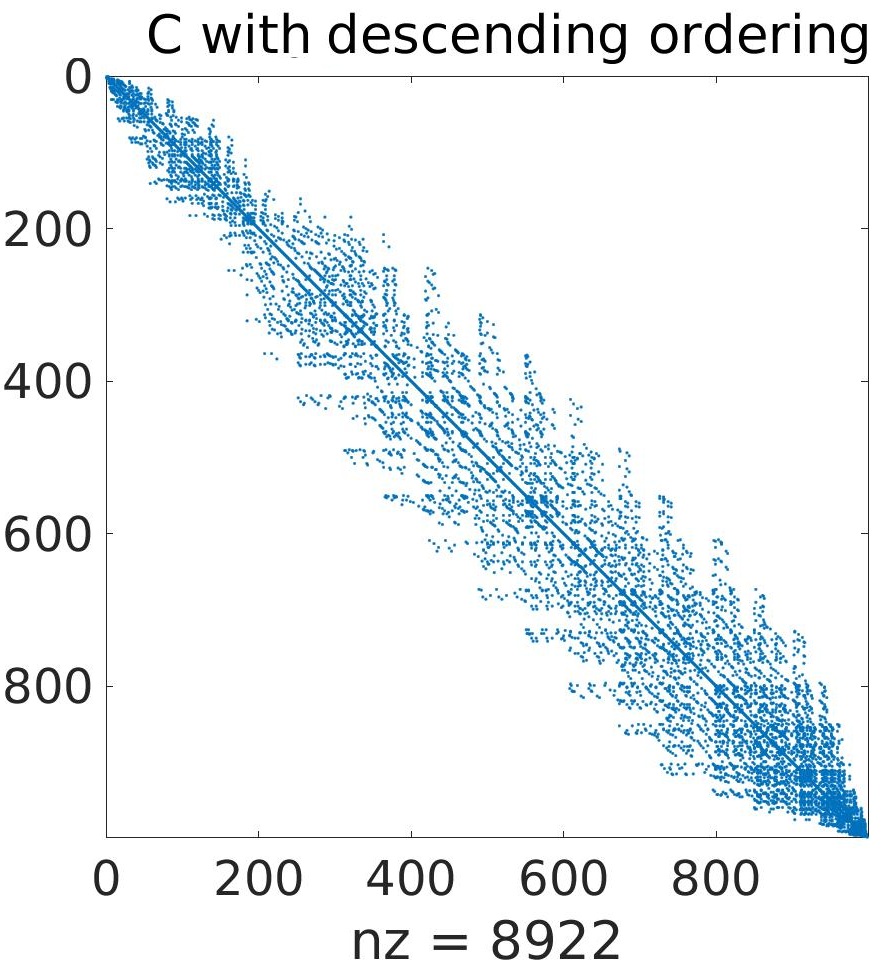} 
	\endminipage		
	\minipage{0.256\textwidth}
	\includegraphics [width=0.89\linewidth]{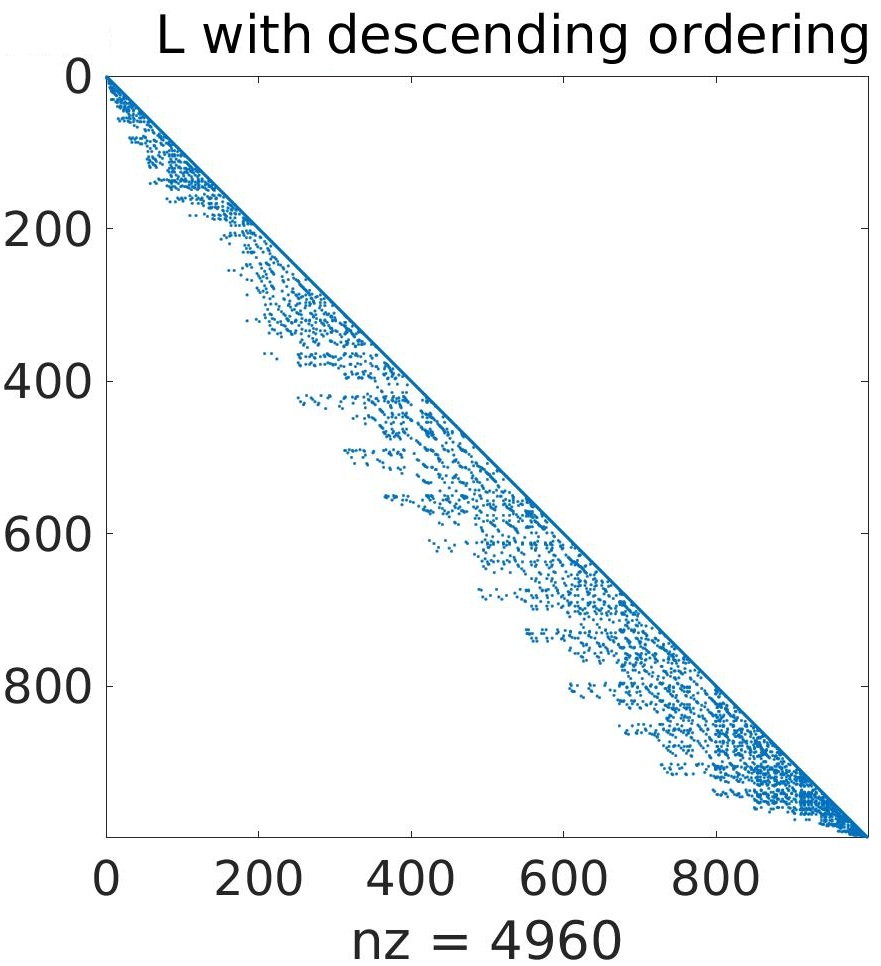}
	\endminipage
	\caption{Nonzero structures of $\mathbf{C}$ and $\mathbf{L}$ from algorithms with and without descending ordering, molecule with $n=1000, r=998$.}
	\label{n1000}
\end{figure}
\begin{figure}[!h]
\centering		
	\minipage{0.26\textwidth}
	\includegraphics[width=0.89\linewidth] {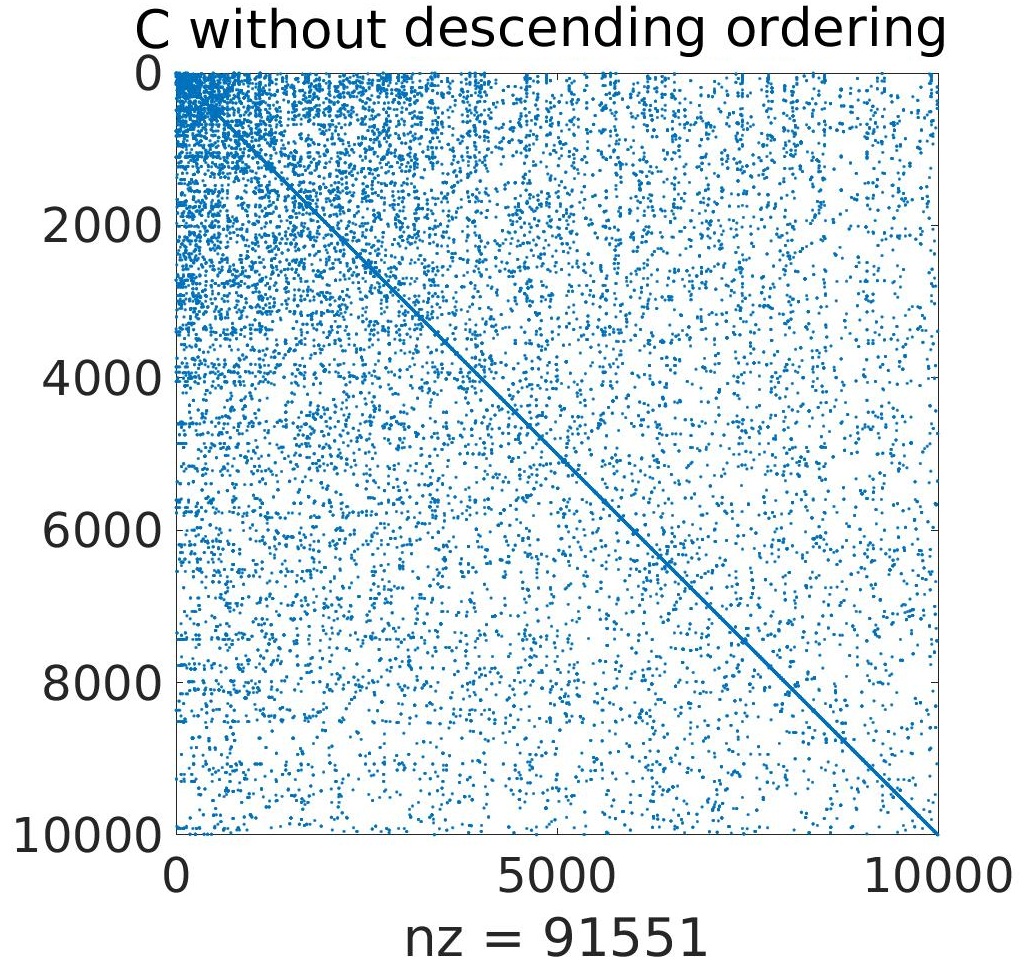} 
	\endminipage		
	\minipage{0.26\textwidth}
	\includegraphics [width=0.89\linewidth]{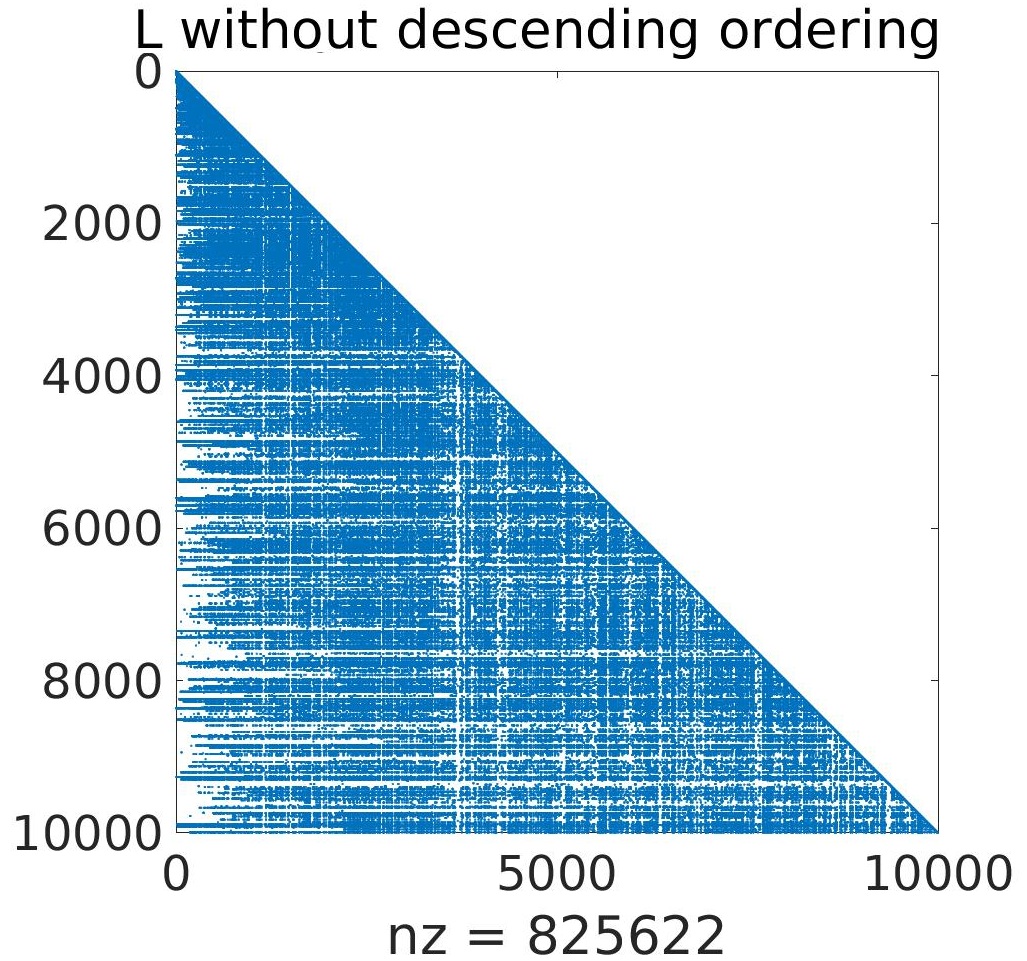}
	\endminipage
	\minipage{0.26\textwidth}
	\includegraphics[width=0.89\linewidth] {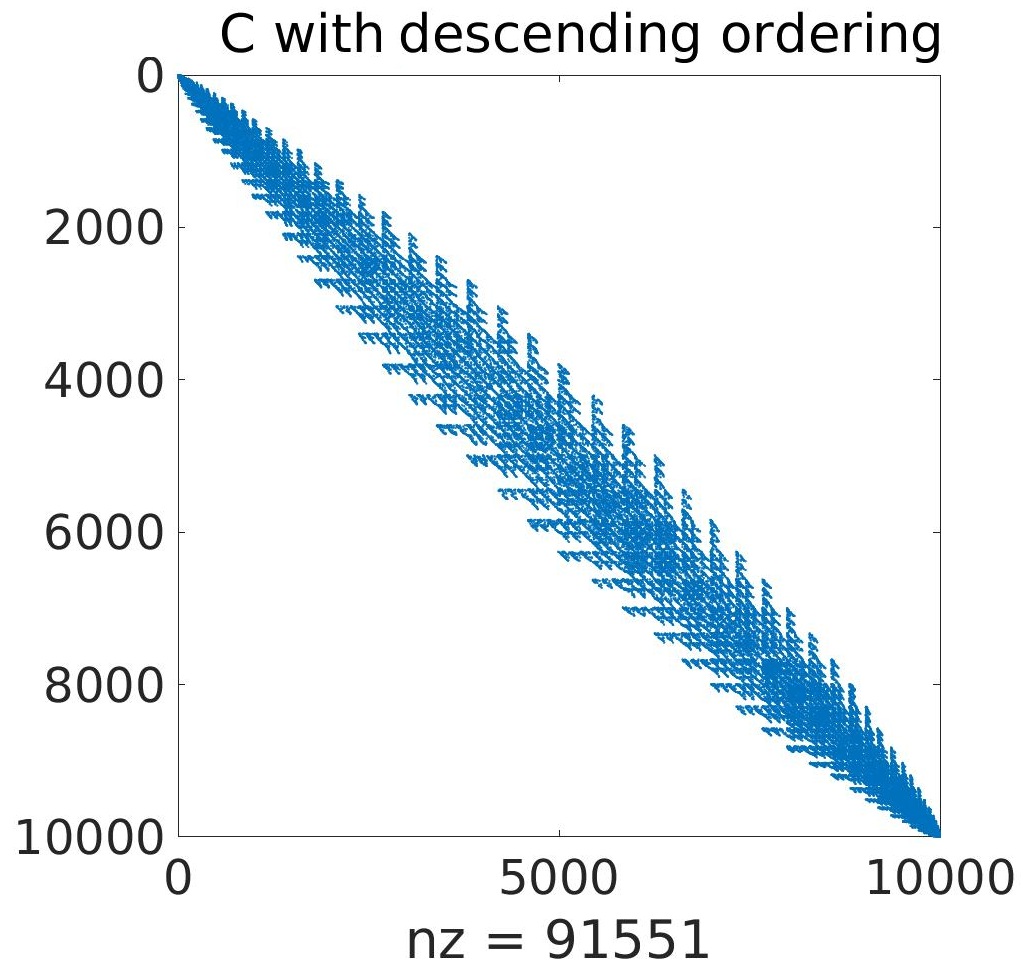} 
	\endminipage		
	\minipage{0.26\textwidth}
	\includegraphics [width=0.89\linewidth]{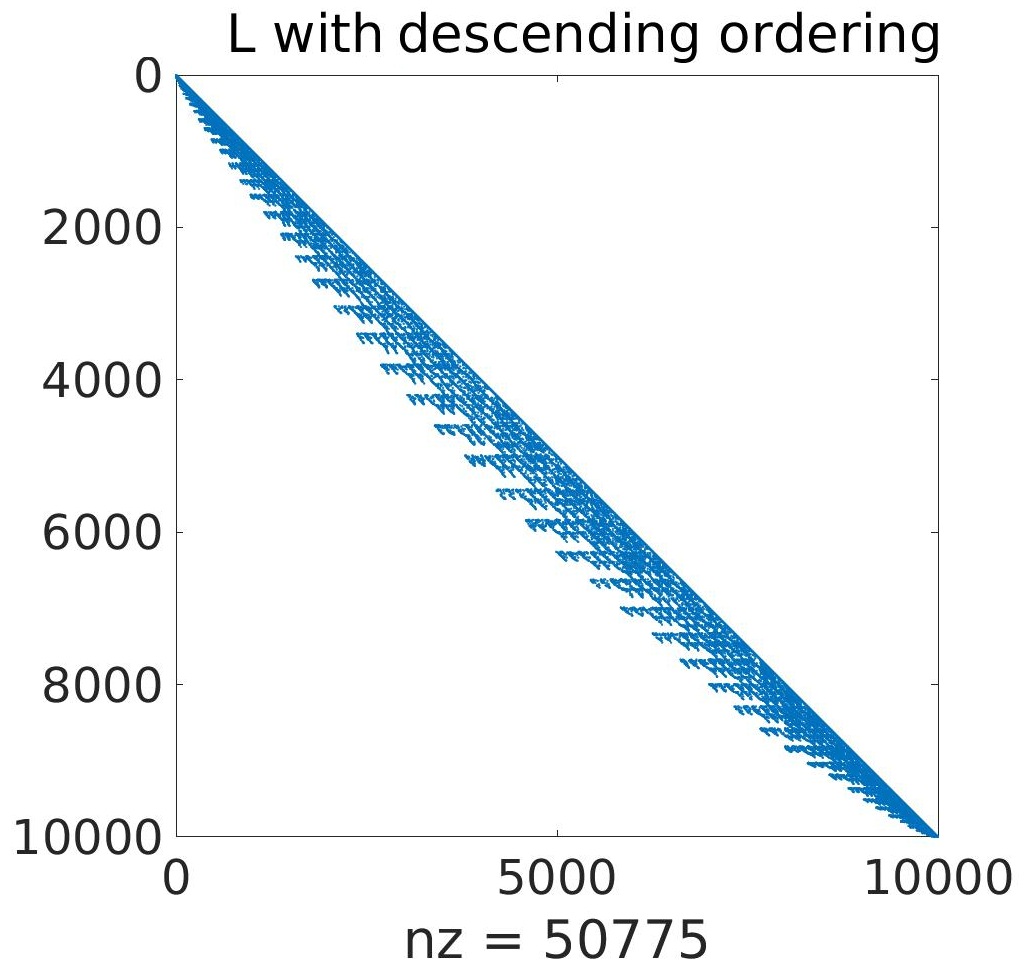}
	\endminipage
	\caption{Nonzero structures of $\mathbf{C}$ and $\mathbf{L}$ from algorithms with and without descending ordering, molecule with $n=10000, r=9999$.}
	\label{n10000}
\end{figure}

In this section, three ``artificial'' molecules with number of atoms $n=100, 1000$ and $10000$ are used to show the no fill-in effectiveness of the distance descending ordering method. Since the sparsity of matrix $\mathbf{C}$ is mainly affected by two factors: the type of the hard variables and the number of branches. To make the comparison be consistent, the hard variables are chosen by randomly fixing one third of $\phi_i$, one third of $\theta_i$, and one third of $b_i$ for all the three molecules. Also the ratio of number of branches to number of atoms is fixed as 0.25. The nonzero structures of all the $\mathbf{C}$ and $\mathbf{L}$ are shown in Figs. (\ref{n100}, \ref{n1000}, \ref{n10000}). Three conclusions can be made by comparing the results:
\begin{enumerate}
\item
	For $\mathbf{C}$ and $\mathbf{L}$ obtained from the distance descending ordering method, the nonzero structure of $\mathbf{L}$ is exactly the same as that of the lower diagonal part of $\mathbf{C}$. The number of nonzero entries in lower diagonal part of $\mathbf{C}$ is $(nz_{\mathbf{C}}-r)/2+r$, which equals to $nz_{\mathbf{L}}$ for all the three cases. Where $nz_{\mathbf{C}}$ and $nz_{\mathbf{L}}$ are numbers of nonzero entries in $\mathbf{C}$ and $\mathbf{L}$.
\item
	The numbers of nonzero entries are the same for both the two matrices $\mathbf{C}$ obtained from algorithms with and without the descending ordering, a permutation can convert one to another.
\item
	With the application of the distance descending ordering method, the number of nonzero entries in $\mathbf{L}$ is roughly linearly scaled with the number of atoms: $nz_{\mathbf{L}}=480$ for $n=100$, $nz_{\mathbf{L}}=4960$ for $n=1000$, and $nz_{\mathbf{L}}=50775$ for $n=10000$. This implies that the solving of the equation  $\mathbf{\dot{q}}=\pmb{\mathcal{M}}^{-1}\mathbf{p}$ will also have this linear scaling. On the other hand, without the descending ordering, the number of nonzero entries in $\mathbf{L}$ grows exponentially with the increase of $n$. 
\end{enumerate}

\begin{figure}[!h]
\centering
	\includegraphics[width=0.60\linewidth]{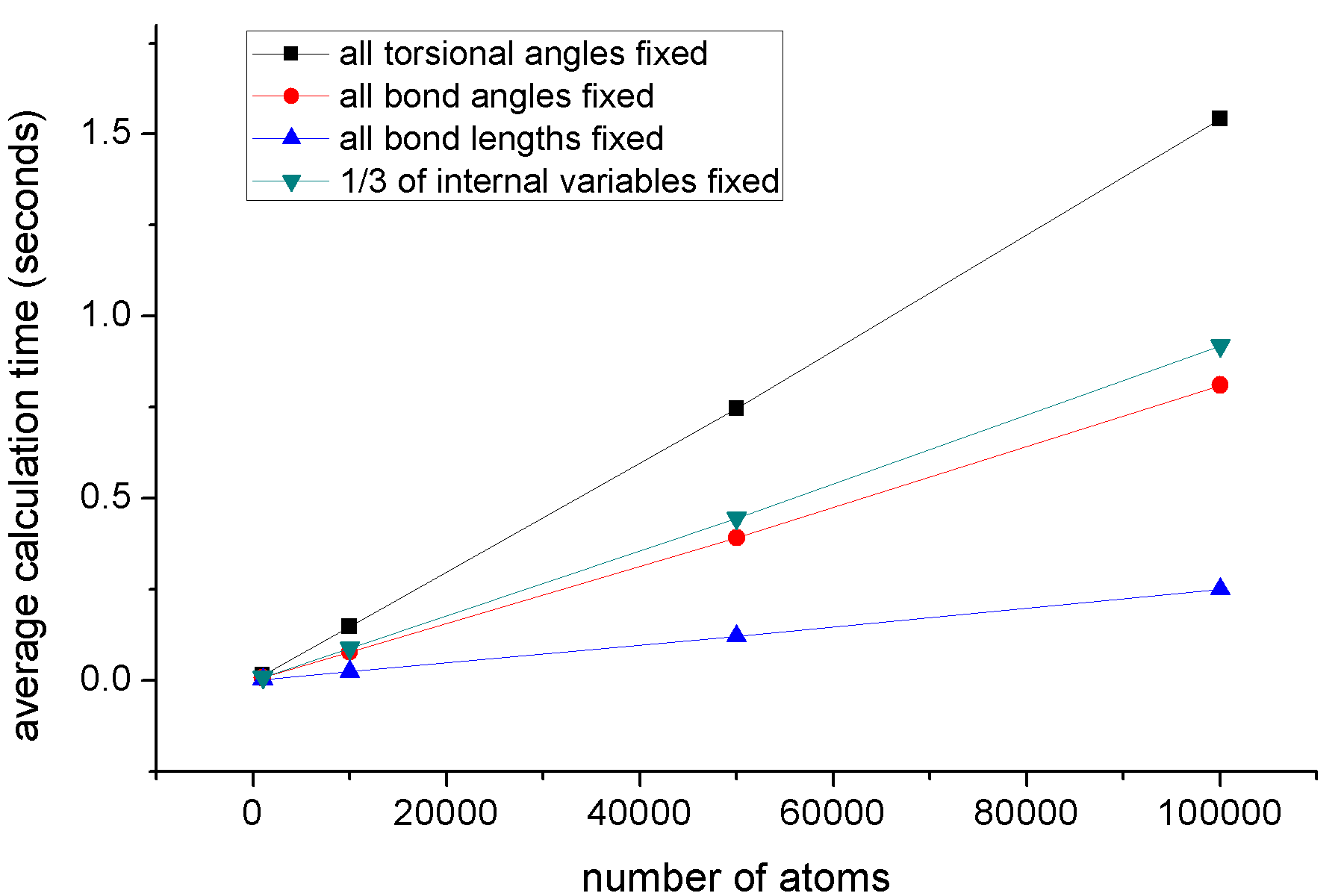}
	\caption{Average calculation time versus number of atoms}
	\label{averagetime}
\end{figure}

The linear computational scaling can also be demonstrated by the average calculation time \replaced{versus the number of atoms}{in solving the equation $\mathbf{\dot{q}}=\pmb{\mathcal{M}}^{-1}\mathbf{p}$}. The number of atoms in a molecule is varied with values $n=1000, 10000, 50000$ and $100000$ to show the $O(n)$ performance. Furthermore, since the type of hard variables have influence on the denseness of $\mathbf{C}$, 4 strategies in choosing the hard variables are used to study this influence. These 4 constraint strategies are:
\begin{enumerate}
\item
	All torsional angles except $\phi_1$ and $\phi_2$ are chosen as hard variables, $r=n-2$.
\item
	All bond angles except $\theta_1$ are chosen as hard variables, $r=n-1$.
\item
	All bond lengths are chosen as hard variables, $r=n$.
\item
	Randomly fixing one third of $\phi_i$, one third of $\theta_i$ and one third of $b_i$, leave $\phi_1, \phi_2, \theta_1$ being free, $r\approx n-3$. 	
\end{enumerate}   
Also, the ratio of number of branches to number of atoms is fixed as 0.25 for all the four molecules. At the beginning of each calculation, the values of internal coordinates are initialized with random numbers, then the equations in appendix \ref{secA1} are adopted to calculate $\frac{\partial \phi_i}{\partial \vec{r}_k}$, $\frac{\partial \theta_i}{\partial \vec{r}_k}$ and $\frac{\partial b_i}{\partial \vec{r}_k}$, and the nonzero entries in $\mathbf{C}$ can be calculated, followed by the Cholesky decomposition from $\mathbf{C}$ to $\mathbf{L}$. Finally, Eqs. (\ref{eqG1} $\sim$ \ref{eqG5}) are solved sequentially. This procedure is repeated 1000 times to give the average calculation time for each $n$ and each constraint strategy. Fig. \ref{averagetime} shows the linear scaling of the distance descending ordering method, as well as the influence of the type of hard variables on the average calculation time. It could be found that all $\phi_i$ fixed constraint strategy has the largest average calculation time, while all $b_i$ fixed constraint strategy has the smallest average calculation time, and the all $\theta_i$ fixed constraint strategy has the medium average calculation time. This agrees with the fact that $\phi_i, \theta_i$ and $b_i$ are affected by $4, 3$ and $2$ atoms, hence $\phi_i$ will correlate with more hard variables than $\theta_i$ and $b_i$. This indicates the more unfixed torsional angles, the faster calculation for real molecular simulations. 

\section{Conclusions}
\label{conclusion}
The Fixman's theorem factorizes the inverse of mass matrix as $\pmb{\mathcal{M}}^{-1}=\mathbf{A-BC}^{-1}\mathbf{B}^{\top}$, from which the equation of motion $\dot{\mathbf{q}}=\pmb{\mathcal{M}}^{-1}\mathbf{p}$ can be decomposed into a series of $O(n)$ calculations, given that $\mathbf{C}$ can be efficiently factorized. For molecules with serial chain structure, the nonzero entries of $\mathbf{C}$ will be clustered about its diagonal, giving linear scaling to the factorization about $\mathbf{C}$. However, for structures with long branches, nonzero entries of $\mathbf{C}$ will no longer clustered about the diagonal, and therefore direct Cholesky decomposition will have $O(n^3)$ computational complexity. Although several methods may be applied in factorizing positive definite sparse matrices, none of them can strictly guarantee no fill-in for all molecule models according to our test. The distance descending ordering method considers the problem from a different perspective: it does not do direct row and column permutations about the matrix $\mathbf{C}$, instead it applies a reordering about the hard variables. The reordering strategy is developed based on the properties of Cholesky decomposition and the geometrical structure of molecules. By applying the distance descending ordering method, the nonzero structure of $\mathbf{L}$ will be exactly the same as that of the lower diagonal part of $\mathbf{C}$. As a result, the $O(n)$ computational efficiency can be remained, no matter the molecule structure has long branches or not.

The distance descending ordering method was proposed with proper proof. The numerical validations have shown its effectiveness. The no fill-in property was demonstrated by using three molecules with number of atoms $n=100, 1000$ and $10000$. The difference between algorithms with and without applying the distance descending ordering method is huge. For results obtained from distance descending ordering method, the number of nonzero entries in  $\mathbf{L}$ is roughly linearly scaled with $n$, this implies the $O(n)$ scaling in solving the equation of motion. However, for algorithm without distance descending ordering method, the order of calculation exponentially grows with the increase of $n$.

The $O(n)$ efficiency of this method has also been demonstrated by the average calculation time versus the molecule's number of atoms.  The authors found that the types of internal coordinates have significant influence on the average calculation time. Fixed torsional angles tend to have more calculation time than fixed bond angles and bond lengths.

\appendix\markboth{Appendix}{Appendix}
\numberwithin{equation}{section}
\section{Calculation of $\frac{\partial \mathbf{q}}{\partial \mathbf{r}}$ and $\frac{\partial \mathbf{c}}{\partial \mathbf{r}}$ }
\label{secA1}
As shown in section \ref{fixmantheorem}, the values of $\frac{\partial \mathbf{q}}{\partial \mathbf{r}}$ and $\frac{\partial \mathbf{c}}{\partial \mathbf{r}}$ are required for the calculation. In other words, we need to calculate $\frac{\partial \phi_i}{\partial \vec{r}_k}, \frac{\partial \theta_i}{\partial \vec{r}_k}$ and $\frac{\partial b_i}{\partial \vec{r}_k}$. The classical book \cite{allen1989computer} has given $\nabla_{\vec{r}_k}\phi_i=\nabla_{\vec{r}_k}\cos\phi_i/(-\sin\phi_i)$ and $\nabla_{\vec{r}_k}\theta_i=\nabla_{\vec{r}_k}\cos\theta_i/(-\sin\theta_i)$. However, this procedure becomes invalid when $\sin \phi_i$ and $\sin \theta_i$ approach zero. Under this condition we need to switch to the alternative approach as $\nabla_{\vec{r}_k}\phi_i=\nabla_{\vec{r}_k}\sin\phi_i/(\cos\phi_i)$ and $\nabla_{\vec{r}_k}\theta_i=\nabla_{\vec{r}_k}\sin\theta_i/(\cos\theta_i)$.

Same as that in \cite{allen1989computer}, two variables are defined to simplify the formulae:
\begin{equation}
\begin{split}
	C_{ij}=C_{ji} & =\vec{b}_i \cdot \vec{b}_j \\
	D_{ij}=D_{ji} & =\left|\vec{b}_i \times \vec{b}_j\right|^2=C_{ii}C_{jj}-C_{ij}^2
\end{split}
\label{CD}
\end{equation}
where $\vec{b}_i=\vec{r}_i-\vec{r}_{i-1}$ is the bond vector. The sines can be obtained as follows:
\begin{equation}
\begin{split}
	\sin \phi_i & =\frac{\vec{b}_{i-1}\times\left(\vec{b}_{i-2}\times\vec{b}_{i-1}\right)}{\left| \vec{b}_{i-1} \right| \left| \vec{b}_{i-2}\times\vec{b}_{i-1} \right|} \cdot 
	\frac{\vec{b}_{i-1}\times\vec{b}_{i}}{\left| \vec{b}_{i-1}\times\vec{b}_{i} \right|} \\
	& =\vec{b}_{i-2}\cdot\left(\vec{b}_{i-1}\times\vec{b}_{i}\right) \sqrt{\frac{C_{i-1 i-1}}{D_{i-2 i-1}D_{i i-1}}} 
\end{split}
\end{equation}
and
\begin{equation}
\begin{split}
	\sin \theta_i & =\frac{\vec{b}_{i}\times\left(\vec{b}_{i-1}\times\vec{b}_{i}\right)}{\left| \vec{b}_{i} \right| \left| \vec{b}_{i-1}\times\vec{b}_{i} \right|} \cdot 
		\frac{\vec{b}_{i-1}}{\left| \vec{b}_{i-1}\right|}  =\sqrt{\frac{D_{i i-1}}{C_{i i} C_{i-1 i-1}}} 
\end{split}
\end{equation}

It is easy to verify that
\begin{equation}
	\begin{cases}
		\frac{\partial \phi_i}{\partial \vec{r}_k}\neq 0 & \text{only if } k=i, i-1, i-2,i-3\\
		\frac{\partial \theta_i}{\partial \vec{r}_k}\neq 0 & \text{only if } k=i, i-1, i-2\\
		\frac{\partial b_i}{\partial \vec{r}_k}\neq 0 & \text{only if } k=i, i-1
	\end{cases}
\label{eqC}
\end{equation}
In the following the nonzero terms of $\nabla_{\vec{r}_k}\sin\phi_i, \nabla_{\vec{r}_k}\sin\theta_i$ and $ \nabla_{\vec{r}_k} b_i$ are given:
\begin{equation}
\begin{split}
	\nabla_{\vec{r}_i}\sin\phi_i = & C_{i-1i-1}^{1/2}\left(D_{i-2 i-1} D_{i-1 i} \right)^{-1/2} \left\{ \left(C_{i i-1} \vec{b}_{i-1} -C_{i-1 i-1} \vec{b}_i  \right) \left[ \vec{b}_{i-2} \cdot \left( \vec{b}_{i-1} \times \vec{b}_{i} \right) \right] / D_{i i-1} \right. \\
	& \left. \vec{b}_{i-2} \times \vec{b}_{i-1}  \right\} \\
	\nabla_{\vec{r}_i}\sin\phi_{i+1} = &  C_{ii}^{1/2}\left(D_{i i-1} D_{i i+1} \right)^{-1/2} \left\{ \vec{b}_{i-2} \cdot \left( \vec{b}_{i-1} \times \vec{b}_{i} \right) \left[ \vec{b}_i/C_{ii} -\left( C_{i-1 i-1} \vec{b}_i -C_{i i-1} \vec{b}_{i-1} \right)/D_{i i-1} \right. \right.\\
	&\left. \left. - \left(C_{ii}+C_{ii+1}\right) \vec{b}_{i+1}/D_{i i+1} + \left(C_{i+1i+1}+C_{ii+1}\right) \vec{b}_{i}/D_{i i+1} \right] + \left(\vec{b}_{i+1} + \vec{b}_{i}\right)\times \vec{b}_{i-1} \right\} \\
	\nabla_{\vec{r}_i}\sin\phi_{i+2} = &  C_{i+1i+1}^{1/2}\left(D_{i i+1} D_{i+1 i+2} \right)^{-1/2} \left\{ \vec{b}_{i} \cdot \left( \vec{b}_{i+1} \times \vec{b}_{i+2} \right) \left[ \left( C_{i+2 i+2} \vec{b}_{i+1} -C_{i+1 i+2} \vec{b}_{i+2} \right)/D_{i+1 i+2} \right. \right.\\
	& \left. + \left(C_{ii}+C_{ii+1}\right) \vec{b}_{i+1}/D_{i i+1} - \left(C_{i+1i+1}+C_{ii+1}\right) \vec{b}_{i}/D_{i i+1} -\vec{b}_{i+1}/C_{i+1 i+1} \right]\\
	&\left. + \left(\vec{b}_{i+1} + \vec{b}_{i}\right)\times \vec{b}_{i+2} \right\} \\
	\nabla_{\vec{r}_i}\sin\phi_{i+3} = & C_{i+2i+2}^{1/2}\left(D_{i+1 i+2} D_{i+2 i+3} \right)^{-1/2} \left[ \vec{b}_{i+1} \cdot \left(\vec{b}_{i+2} \times \vec{b}_{i+3} \right) \left(C_{i+2 i+2}\vec{b}_{i+1} -C_{i+1 i+2} \vec{b}_{i+2} \right)/D_{i+1i+2} \right. \\
	& \left. +\vec{b}_{i+3}\times \vec{b}_{i+2}  \right] 
\end{split}
\end{equation}
and
\begin{equation}
\begin{split}
	\nabla_{\vec{r}_i}\sin\theta_i =& D_{i-1i}^{1/2}\left(C_{i-1 i-1} C_{i i} \right)^{-1/2}\left[ \left(C_{i-1i-1}/D_{i-1 i}-1/C_{ii}\right) \vec{b}_i -C_{i-1i}/D_{i-1i}\vec{b}_{i-1} \right]\\
	\nabla_{\vec{r}_i}\sin\theta_{i+1} =& D_{ii+1}^{1/2}\left(C_{i i} C_{i+1 i+1} \right)^{-1/2} \left\{ \left[\left(C_{i+1 i+1} +C_{i i+1}\right) \vec{b}_i- \left( C_{ii}+C_{ii+1} \right)\vec{b}_{i+1} \right]/D_{ii+1} \right. \\
	& \left. +\vec{b}_{i+1}/C_{i+1i+1} -\vec{b}_i/C_{ii} \right\}\\
	\nabla_{\vec{r}_i}\sin\theta_{i+2} =& D_{i+1i+2}^{1/2}\left(C_{i+1 i+1} C_{i+2 i+2} \right)^{-1/2} \left[ \left(C_{i+1i+2}\vec{b}_{i+2} -C_{i+2i+2}\vec{b}_{i+1} \right)/D_{i+1i+2}  +\vec{b}_{i+1}/C_{i+1i+1} \right]
\end{split}
\end{equation}
also,
\begin{equation}
\begin{split}
	\nabla_{\vec{r}_i}b_i = & \vec{b}_i/b_i \\
	\nabla_{\vec{r}_i}b_i = & -\vec{b}_{i+1}/b_{i+1}
\end{split}
\end{equation}

\bibliographystyle{unsrt}
\bibliography{distancedescendingorderingmethod}
\end{document}